\documentstyle[preprint,aps]{revtex}
\tighten
\draft
\preprint{\vbox{ \hfill IU/NTC 96-18 \\
   \null \hfill hep-ph/9701214 }}
\begin{document}
\title{
Macroscopic Parity Violation and Supernova Asymmetries}
\author{C. J. Horowitz~\footnote{E-mail:~Charlie@IUCF.indiana.edu}}
\address{Nuclear Theory Center and Department of Physics, \\
	 Indiana University, Bloomington, IN 47405, USA}
\author{J. Piekarewicz~\footnote{E-mail:~jorgep@scri.fsu.edu}}
\address{Supercomputer Computations Research Institute, \\
      Florida State University, Tallahassee, FL 32306}

\date{\today}
\maketitle
\begin{abstract}
Core collapse supernovae are dominated by weakly interacting neutrinos.  This
provides a unique opportunity for macroscopic parity violation.  We speculate
that parity violation in a strong magnetic field can lead to an asymmetry in
the explosion and a recoil of the newly formed neutron star.  We estimate
the size of this asymmetry from neutrino polarized neutron elastic scattering,
polarized electron capture and neutrino-nucleus elastic scattering in a
(partially) polarized electron gas.
\end{abstract}
\narrowtext
\section{Introduction}
\label{introduction}

Core collapse supernovae emit approximately $10^{58}$ neutrinos. Apart
from gravity, these neutrinos only have weak interactions---whose
hallmark is parity violation. Can this lead to {\it macroscopic} parity 
violation?  The important role played by weak interactions makes 
supernovae (or related phenomena) unique among all macroscopic systems 
in the present universe. No other systems have this potential for large 
scale parity violation.

One parity violating observable is a correlation 
${\bf  k_\nu}\cdot {\bf B}$ between the neutrino flux in the direction
${\bf  k_\nu}$ and the direction of the magnetic field ${\bf B}$.  
As a result, the entire neutron star, formed in the explosion, could 
recoil. This can be at a significant velocity because of the tremendous
momentum of the neutrinos (see below).

Indeed, the observed velocities of neutron stars are large: with three 
dimensional galactic velocities of the order of 500 km/s \cite{vel}.  
Perhaps the simplest explanation for such large velocities is that 
neutron stars receive a significant kick at birth from an asymmetric 
supernova explosion. Note that this asymmetry need not be solely from 
parity violation. For example, one could have a preexisting asymmetry 
in the collapsing stellar core as speculated by Burrows and 
Hayes~\cite{burrows}.  These authors also discuss how an asymmetry
may impact a detectable gravitational radiation signal.

In any case, there is a simple relation between the final velocity 
of a neutron star and the asymmetry of the neutrinos $A_\nu$ and other 
matter $A_m$ ejected.  Note, a supernova ejects the mantle and outer envelope
comprising some 90 percent of the original star's mass.  The binding 
energy per nucleon of a neutron star
is about 100 MeV/n. Thus, neutrinos carry off 100 MeV/c of momentum per 
nucleon of the neutron star (which then has a gravitational mass of 
the order of 839 MeV/n). In contrast, other matter only carries off 
about one percent of the binding energy (1 MeV/n). However, the mass of 
this material is about 10 times larger than that of the neutron star, 
so that its momentum ($p_{\rm m}\approx [2\cdot 1{\rm MeV}\cdot 
10\cdot 939 {\rm MeV}]^{1/2}\approx 140$~MeV/c) is comparable to that 
of the neutrinos.\footnote{One can ask why do supernovae explode so 
that the momentum in matter is comparable to that in neutrinos?}

The recoil velocity $v$ of a neutron star is approximately,
\begin{equation}
v/c \approx 0.1 ( A_\nu + A_m),
\end{equation}  
with $c$ the speed of light and the asymmetry of the neutrinos is 
\begin{equation} 
A_\nu={|\sum_i {\bf p_i}|\over \sum_i |{\bf p_i}|},
\end{equation} 
where the sum runs over all neutrinos emitted.  There is a 
similar expression for the asymmetry of the other matter $A_m$.  
{\it To reproduce observed velocities of the order of $10^{-3}c$\ 
one needs $A_\nu$---and/or $A_m$---to be about one percent.}

Since matter and neutrinos are coupled, one expects both $A_\nu$ 
and $A_m$ to be nonzero, i.e., an original asymmetry in one will 
produce an asymmetry in the other.  Thus, the interesting question 
is which came first (the chicken or the egg) $A_\nu$ or $A_m$?

In this paper we speculate that the original $A_\nu$ may stem from 
parity violation in a strong magnetic field.  We assume conventional 
weak interactions for the neutrinos.  Others have considered
nonstandard neutrino magnetic moments \cite{magmom} or matter 
enhanced neutrino oscillations \cite{osc}.  These could enhance 
the effect.  However, one does not require new interactions for a 
nonzero asymmetry.  Thus, our speculation only depends on the 
existence of a strong magnetic field.

A number of Pulsars are thought to have magnetic fields around 
$10^{12}$ Gauss.  However, the dipole field inside a supernova at 
early times could be stronger, perhaps $10^{14}$ Gauss~\cite{magnetic}.  
Finally, it is possible that there are very strong {\it non-dipole} 
fields of the order of $10^{16}$ Gauss~\cite{strongmag}.  For example,
differential rotation could wind up the ${\bf B}$ field into a very 
strong torus configuration.  In section II we calculate the neutrino 
asymmetry that these fields induce because of several parity violating
hadronic and electron reactions.  We discuss possible enhancements of 
the asymmetry and conclude in section III.

We end this section with a speculative note regarding macroscopic 
parity violation and the origin of life.  Cline suggests~\cite{cline}
that polarized leptons from a nearby supernova could influence homochiral
organic molecule formation.  This is an interesting but clearly very
speculative atempt to explain why Terrestrial life uses almost 
exclusively L-amino-acid enantiomers (``left handed'' mirror image molecules).
Although not directly related, we speculate that supernovae produce another 
macroscopic parity violating effect.  The hypothesis that parity violation 
leads to an asymmetry and recoil may be more directly observed and tested.

\section{Microscopic Parity Violating Reactions}
\label{microscopic}

There are many possible sources of parity violation in a strong
magnetic field.  One would expect weak interactions involving
electrons to be the most important because of the large electron
magnetic moment.  However, a large Fermi momentum can make it harder
to polarize electrons.  Furthermore, relativistic effects reduces the
effective magnetic moment.  Finally, interactions with nucleons often
dominate the neutrino opacity rather than interactions with electrons.
Therefore, it is possible for nucleon reactions to compete with
electrons.

In this section we consider a number of reactions.  Perhaps the
simplest is neutrino elastic scattering from polarized neutrons.  
The longitudinal asymmetry for this process is very large,
\begin{equation}
|A_l | = {2 g_v g_a\over g_v^2 + 3 g_a^2} \ \ \approx 0.46,
\end{equation}
with $g_v=-1$ and $g_a=-1.26$.  The polarization of neutrons in a
magnetic field $B$ is,
\begin{equation}
P_e \approx eB/M_n kT \;,
\end{equation}
where $M_n$ is the neutron's mass, $T$ the temperature and we have
neglected possible enhancements from the spin dependence of the strong
interactions (more on this below).  Elastic neutron scattering makes a
significant contribution to the neutrino opacity.  Therefore, this
will produce an asymmetry in the neutrino flux of the order of
$A_\nu \approx P_e A_l$, which at a temperature near 3 MeV is,
\begin{equation}
|A_\nu| \approx 1.3 \times 10^{-4} B_{14} \;,
\label{neutron}
\end{equation}
with $B_{14}$ the magnetic field in units of $10^{14}$\ Gauss.  This
asymmetry, of the order of $eB/M_nkT$, is somewhat small compared 
to one percent. However, this simple mechanism is clearly present and
provides a benchmark to compare other reactions.

Vilenkin~\cite{rusNES} estimated the asymmetry of neutrino electron
scattering (NES) while Bezchastnov and Haensel~\cite{NES} calculated
it in detail; it is of the order of $eB/k_F^2$, where $k_F\approx 20$ 
MeV is the electron Fermi momentum.  Here, one power of $k_F^{-1}$ 
represents the relativistic electron magnetic moment and the second power
describes the difficulty of polarizing a degenerate Fermi gas.  The
NES asymmetry is about 10 times the neutron polarization\footnote{Not
1000 times as would be expected from the ratio of magnetic moments in
nonrelativistic non-degenerate gases}.  However, NES makes less then a
10 percent contribution to the total neutrino opacity (indeed it is
probably less then one percent).  Therefore the NES contribution to
$A_\nu$ is probably smaller than Eq. (\ref{neutron}).  Note, NES does
change the neutrino energy and could introduce an asymmetry in the
spectrum.  One should investigate this effect; however, it is
probably small.  For example Haxton and Bruenn made large changes in
the energy loss cross section and found only small effects on the
dynamics~\cite{bruenn}.

We now consider neutrino-nucleus elastic scattering in a polarized
electron gas.  Nuclear scattering dominates the opacity, as long as
many nuclei are present.  Furthermore, this cross section is impacted
by screening from the dense electron gas~\cite{elec,ruselec}.
Polarization of the electron gas by the $B$ field will introduce an
asymmetry in the nuclear cross section even for spin zero nuclei.
Furthermore, the asymmetry will depend on the large electron magnetic
moment.  Therefore this process could produce a significant neutrino
asymmetry.

We are not aware of an exact calculation.  Instead, we present a
simple phase space estimate assuming the magnetic field slightly
polarizes the electron gas.  We are interested in the interference of
the electron particle-hole excitation shown in Fig. 1B with the direct
neutrino-nucleus scattering of Fig. 1A.  Parity violation stems from
the weak axial charge of the electron interfering with the weak vector
charge of the nucleus.  The electron particle-hole loop is described
by (see ref.~\cite{elec})
\begin{equation}
\Pi^{va}_{\mu\nu}(q) = -i \int {d^4p\over (2\pi)^4} {\rm 
tr }[ \gamma_5\gamma_\mu G(p)\gamma_\nu G(p+q)],
\label{pol}
\end{equation}
here $q$ is the four-momentum transferred to the nucleus and the 
electron propagator~\cite{SW} $G=G_F+G_D$ has the usual Feynman piece
$G_F(p)=[\rlap/{p} - m +i \epsilon]^{-1}$ and a density dependent part
$G_D$ which describes the occupied states in the (slightly polarized)
Fermi sea,
\begin{equation}
G_D(p)= (\rlap/{p} + m ) {i\pi \over \sqrt{p^2+m^2}} 
\delta(p_0 -\sqrt{p^2+m^2})
\left\{ n_+\left({1+\gamma_5\rlap/{s}\over 2}\right) + n_-
\left( {1-\gamma_5\rlap/{s}\over 2}\right)\right\}.
\end{equation}
Here $n_+$ and $n_-$ are the occupations of the
Fermi gas states polarized along or against $\hat s$; for an electron
at rest, the relativistic spin vector $s$ is chosen in the direction 
of ${\bf B}$.  We assume a dispersion relation,
\begin{equation}
 \epsilon_\pm^2 \approx p^2+m^2 \pm eB,
\end{equation}
which describes the interaction of the spin magnetic moment with $B$.
We assume that the orbital contributions are small since in the weak 
field limit a very large number of Landau levels will be filled. In 
any event, it is hoped that this simple approximation provides an 
order of magnitude estimate. We evaluate Eq.~(\ref{pol}) in the
extreme relativistic limit $m\rightarrow 0$, for zero energy transfer 
$q_0=0$, and assume weak fields $eB\ll k_F^2$. Moreover, at low 
temperatures one has $n_\pm(p)=\theta( k_F-\epsilon_\pm)$, with 
$k_F$\ the Fermi energy. In a coordinate system with ${\bf q}$ along
$\hat 3$ and $\hat s$ in the 1-3 plane one has,
\begin{equation}
\Pi^{va}_{10}(q) = {{\rm sin} \alpha\over 4\pi^2} eB 
\left\{ {k_F\over q}\left( 1-{q^2\over 4k_F^2}\right)
{\rm ln}\Big|{2k_F+q\over 2k_F-q}\Big| + 1 \right\} \;,
\end{equation}
with $\alpha$ the angle between ${\bf B}$ and ${\bf q}$.  
In the limit of $q\ll k_F$ one has simply,
\begin{equation}
\Pi^{va}_{10}\approx {\rm sin} \alpha {eB \over 2\pi^2}.
\end{equation}
The density independence of this result arises from a cancellation.
The polarization of a relativistic electron gas decreases with density
as $eB/k_F^2$\ while the density of states for particle-hole
excitations goes like $k_F^2$. The asymmetry in the neutrino-nucleus
elastic cross section is,
\begin{equation}
|A| \approx {2Ze^2c_a\over Cq^2}\Pi^{va}_{01}.
\label{A}
\end{equation}
Here $c_a=\pm1/2$\ is the weak axial charge of the electron 
($+$ for $\nu_\mu$, $\nu_\tau$ and $-$ for $\nu_e$), 
$C=Z(1/2-2{\rm sin}^2\theta_W) -N/2$\ is the weak vector charge 
of a nucleus with charge $Z$ and neutron number $N$ and the momentum 
transfer is $q\approx E_\nu\approx 10 MeV$. As long as the opacity 
is dominated by neutrino nucleus scattering this will produce a 
neutrino asymmetry of about,
\begin{equation}
|A_\nu| \approx 5 \times 10^{-5} B_{14},
\end{equation}
for nuclei near $^{56}$Fe.  This somewhat small number arises from the
factor of $e^2=4\pi\alpha$\ in Eq.~(\ref{A}).  Nevertheless, because
of the large electron magnetic moment, $A_\nu$ is comparable to that
from polarized neutrons in Eq. (\ref{neutron}).

Finally, we make a simple phase space estimate for the asymmetry from
polarized electron capture on protons.  For simplicity, we neglect the
electron mass and the neutron-proton mass difference.  The rate for
left handed electrons moving in the direction $\theta_e$ to be
captured and produce a neutrino moving in the direction $\theta_\nu$,
$\phi_\nu$ into $d\Omega$ is,
\begin{equation}
{dR\over d\Omega } \propto \int\ \bigl[ 3g_a^2 + g_v^2 + (g_v^2-g_a^2)
{\rm cos} \theta_{\nu e}\bigr] E_\nu^2 n_{k\downarrow} (\theta_e)
d^3k.
\end{equation}
Here ${\bf k}$ is the momentum of the initial electron and 
$n_{k\downarrow}$ is the occupation of left handed spin states 
(for an electron moving at an angle $\theta_e$ with respect to 
${\bf B}$). Finally, $\theta_{\nu e}$ is the scattering angle of 
the neutrino with respect to the initial electron's direction.

The angular distribution is important. Electron capture on a proton
involves both Fermi transitions with a forward peaked angular
distribution $g_v^2(1+{\rm cos}\theta_{\nu e})$ and backward peaked
Gamow Teller transitions involving $g_a^2(3-{\rm cos}\theta_{\nu e})$.
The sum is only weakly backward peaked ($g_a^2 > g_v^2$).  A flat
angular distribution will not produce a neutrino asymmetry because the
neutrino ``forgets'' the initial electron direction, i.e., neutrinos
are emitted isotropically, independent of any asymmetry in the initial
electron direction.  Therefore only the cos$\theta_{\nu e}$ term will
contribute and it involves the small factor $g_v^2-g_a^2$.

The occupation of left handed electrons along ${\bf k}$ can be 
decomposed into spin up 
$n_+$ and down $n_-$ occupations along ${\bf B}$:
\begin{equation}
n_{k\downarrow}(\theta_e) = ({1+{\rm cos}\theta_e\over 2}) n_- +
({1-{\rm cos}\theta_e\over 2}) n_+
\end{equation}
As above we use $E_\nu^2=E_e^2\approx k^2 \pm eB$, $n_\pm=
\theta[k_F^2-(k^2\pm eB)]$ so that,
\begin{equation}
{dR\over d\Omega} \propto \int d^3k \left[ 1 + {g_v^2-g_a^2\over
3g_a^2+g_v^2} {\rm cos}\theta_{\nu e} \right] \left\{ \left({1+{\rm
cos}\theta_e\over 2}\right)(k^2-eB) n_- + \left({1-{\rm
cos}\theta_e\over 2}\right)(k^2+eB) n_+ \right\}.
\end{equation}
This gives,
\begin{equation}
{dR\over d\Omega} \propto 1 + \left({g_v^2-g_a^2\over
3g_a^2+g_v^2} \right)
\left({5eB\over 12k_F^2}\right) {\rm cos}\theta_\nu,
\end{equation}
and produces a neutrino asymmetry of,
\begin{equation}
A_\nu \approx \left({g_v^2-g_a^2\over 3g_a^2+g_v^2}\right) {5eB\over
12k_F^2}.
\end{equation}
We estimate the contribution from the emission of neutrinos near the
neutrinosphere at a density of the order of $10^{11}$ g/cm$^3$ and at
an electron fraction (number of electrons per baryon) of the order 
of $0.1$.  Here $k_F\approx 10$ MeV and
\begin{equation}
|A_\nu| \approx 2.4 \times 10^{-4} B_{14}.
\end{equation}
This asymmetry is somewhat larger than the others that we have 
calculated. Yet, it has been reduced by a factor of about 30 
because of the nearly flat angular distribution.  If it were not 
for this reduction, this reaction could produce a one percent 
asymmetry for $B_{14}$ near unity.  Note that one should check 
our phase space estimate with an exact calculation.  If the 
${\bf B}$ field modifies the angular distribution it could 
substantially increase the asymmetry.

We illustrate the sign of the electron capture asymmetry in Fig. (2).
A vertical ${\bf B}$ field polarizes electrons with spin down.  
As a result, only electrons moving up are left handed and can be 
captured.  However these electrons produce neutrinos moving down 
because the angular distribution is (weakly) backward peaked.  
Thus the star recoils in the direction of the ${\bf B}$ field (up).

\section{Discussion and Conclusions}

In this section we discuss possible enhancements of the asymmetry and
conclude.  Spin dependent strong interactions can enhance the magnetic
response of neutron rich matter.  This could increase the asymmetry
from polarized neutrons.  For example, Kutschera and
Wojcik~\cite{ferro} speculate that a ferromagnetic state could form
because of spin-dependent neutron-proton interactions.  Indeed the
asymmetry could still be increased by a significant amount, even if a
ferromagnetic state did not form.  However, this enhancement is
unlikely to be an order of magnitude or more unless one is very close
to a ferromagnetic transition---which we think unlikely.  Therefore,
we do not expect very large enhancements from the strong interactions.

Alternatively, the nonlinear dynamics in a supernova could enhance
small microscopic asymmetries.  Indeed, the explosion energy may be
very sensitive to the neutrino heating rate.  Consider the extreme
limit of a heating rate that just fails to produce an explosion.  A
small asymmetry in the neutrino flux will produce a large asymmetry in
the ejected mater by restarting the shock wave on only one side of the
supernova.\footnote{The sensitivity may be somewhat reduced by
requiring the shock to reproduce observed (relatively large) explosion
energies.}  Perhaps a $10^{-3}$\ asymmetry in the neutrino flux can
lead to a $10^{-2}$\ asymmetry in the ejected matter.  This should be
explored with simulations incorporating a small microscopic asymmetry.

We have considered a number of electron reactions which yield somewhat
small asymmetries for perhaps accidental reasons.  For example,
polarized electron capture produces a reduced asymmetry because of a
broad angular distribution.  The natural size of an asymmetry is of
the order of $eB/k_F^2$ for degenerate conditions or $eB/E_\nu^2
\approx 0.6 \times 10^{-2}\ B_{14}$\ for
nondegenerate conditions (with $E_\nu \approx 10$ MeV the neutrino
energy).  This is near one percent for $B_{14}$ near unity. Thus, one
should search for other electron reactions with large asymmetries.

One possibility is $\nu\bar \nu \rightarrow e^+e^-$\ which will be
examined in future work. Note that Kuznetsov and Mikheev~\cite{kuz} 
considered the related $\nu\rightarrow \nu e^+e^-$ in a strong field.
They find a relatively small asymmetry in $A_m$\ of the order of
$10^{-5}$\ $B_{14}$. Perhaps this is because $B$ is necessary to 
produce, both, the asymmetry and to make the process kinematically 
allowed.  Thus we expect $\nu \bar\nu\rightarrow e^+e^-$ to be more 
important because it is allowed even in the absence of $B$.

Supernovae, because they are dominated by weakly interacting
neutrinos, provide a unique opportunity for macroscopic parity
violation.  One parity violating observable is a correlation between
the neutrino flux and the magnetic field directions. We have examined
possible asymmetries in a supernova from known parity violating weak
interactions in strong magnetic fields.  To explain the large
velocities of neutron stars one needs an asymmetry in the radiated
neutrinos {\it and or} the ejected matter of the order of one
percent. We have looked at a number of hadronic and electron reactions
that give asymmetries of the order of a few times $10^{-4}\ B_{14}$.
These are somewhat small to directly produce the recoil velocity from
a dipole magnetic field.  However, they only involve known weak
interactions and, thus, are clearly present and provide a benchmark to
compare with more speculative possibilities.  Furthermore, these
asymmetries are important if supernovae involve very strong non-dipole
magnetic fields (possibly as high as $10^{16}$ Gauss).  One should
continue to search for new reactions and or enhancements since the
natural size of electron reaction asymmetries $eB/E_\nu^2\approx 
0.6 \times 10^{-2}\ B_{14}$\ could
produce a large enough effect.

\acknowledgments
This work was supported by the DOE under grant numbers
DE-FG02-87ER40365, DE-FC05-85ER250000 and DE-FG05-92ER40750.

\vbox to 4.in{\vss\hbox to 8in{\hss
{\includegraphics{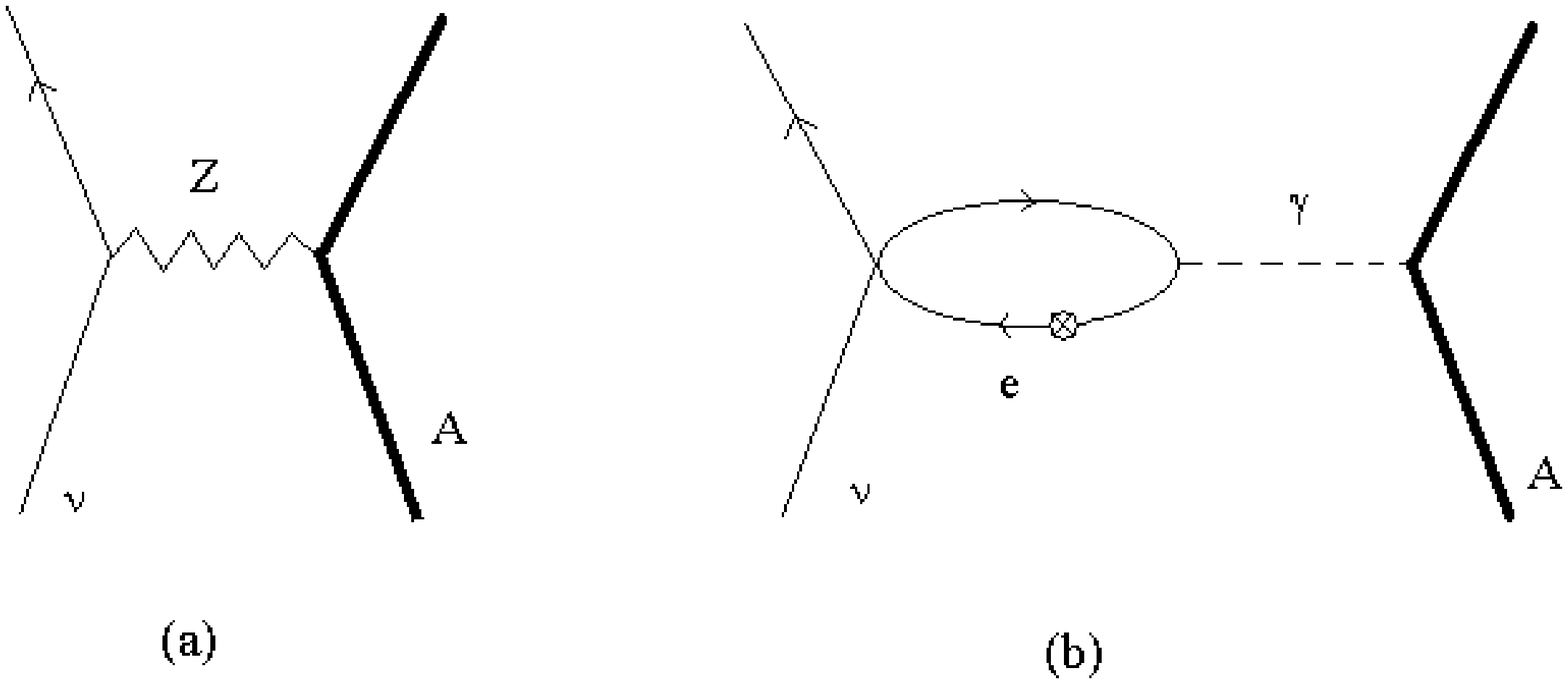}}\hss}}
\nobreak
{\noindent\narrower{{\bf FIG.~1}.   Neutrino-nucleus
elastic scattering in a polarized electron gas.  The direct
neutrino-nucleus diagram $(a)$ interferes with the electron
screening diagram $(b)$ where the neutrino first excites an
electron particle-hole state.  This introduces a parity violating
asymmetry in the angular distribution because the electron gas is
(slightly) polarized by a magnetic field as indicated by the $X$.}}
\bigskip

\vbox to 4.in{\vss\hbox to 8in{\hss
{\includegraphics{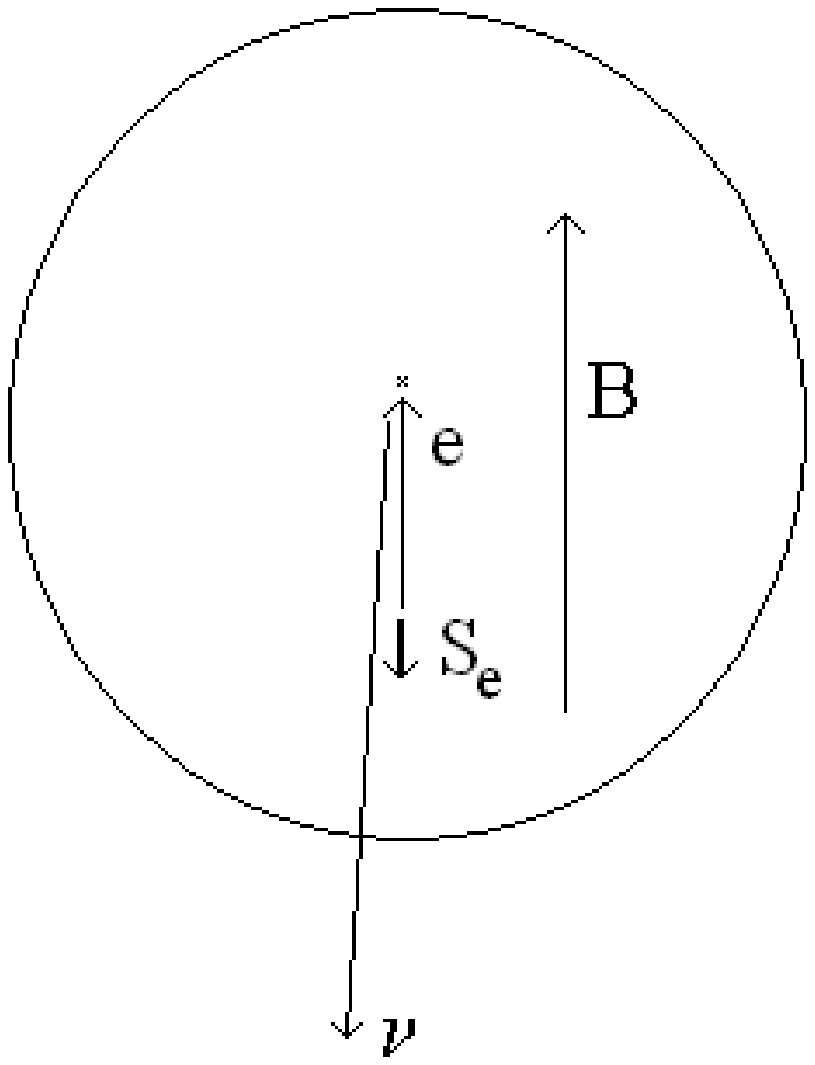}}\hss}}
\nobreak
{\noindent\narrower{{\bf FIG.~2}.  Recoil of a neutron star from polarized
electron capture.  A vertical $B$ field polarizes electrons with spin down 
(indicated by the arrow $S_e$).  These electrons when moving up are left 
handed and can be captured.  However, the electron capture angular 
distribution is backward peaked so that neutrinos will tend to be emitted
down.  As a result the neutron star will recoil up, in the direction of
the magnetic field.}}

\end{document}